\documentclass[twocolumn,prb]{revtex4}
\usepackage{amsfonts}
\usepackage[T1]{fontenc}
\usepackage{amsmath,amsbsy,amssymb,graphicx}
\usepackage{times}
\usepackage{color}

\let\mathbf=\boldsymbol

\begin{document}

\title{Triplet Fermions and Dirac Fermions in Borophene}
\author{Motohiko Ezawa}
\affiliation{Department of Applied Physics, University of Tokyo, Hongo 7-3-1, 113-8656,
Japan}

\begin{abstract}
Borophene is a monolayer materials made of boron. A perfect planar
boropehene called $\beta_{12}$ borophene has Dirac cones and they are well
reproduced by a tight-binding model according to recent experimental and
first-principles calculation results. We explicitly derive a Dirac theory 
for them. Dirac cones are gapless when the inversion
symmetry exists, while they are gapped when it is broken. In addition,
three-band touching points emerge 
together with pseudospin
triplet fermions when all transfer energy is equal and all on-site energy
is equal. The three-band touching is slightly resolved otherwise. 
We construct effective three-band theories for triplet
fermions. We also study the edge states of borophene nanoribbons, which
show various behaviors depending on the way of edge
terminations.
\end{abstract}

\maketitle

\section{Introduction}

Monolayer material science is one of the most active fields of condensed
matter physics in this decade. It has begun with graphene\cite{NetoRev} and
been extended to the group IV monolayer materials including silicene\cite%
{LiuPRB,EzawaQAHE,LTao}, germanene\cite{Davi,Der,GLi} and stanene\cite%
{Stanene}. Furthermore, experimental success of phosphorene\cite%
{Phos1,Phos2,Phos3} has opened a field of the group V monolayer materials
including arsenene\cite{Arsenene} and antimonene\cite{Antimonene}. A search
for new monolayer materials is extended to the group III monolayer materials
including borophene and aluminene\cite{Alumi}. Especially, several types of
borophene are proposed by first-principles calculation\cite%
{HTang,XWu,Penev,HLiu,ZZhang2,YLiu}. Recently, several types of borophene is
synthesized on Ag(111)\cite{Man,ZZhang,BFeng,BFeng2}.

In particular, $\beta_{12}$ borophene experimentally manufactured on the
silver surface\cite{Boro} is quite interesting. Dirac fermions are clearly
observed by the ARPES experiments as well as by first-principles
calculation. The band structure is well reproduced by a tight-binding model,
where it is enough to take into account only the $p_{z}$ orbitals due to its
perfect planar structure. The unit cell contains five atoms as in Fig.\ref%
{FigLattice}(a). The lattice has a perfectly flat structure as in Fig.\ref%
{FigLattice}(b). It can be constructed by adding atoms indicated in yellow
into the honeycomb lattice.

In this paper we study the band structure of $\beta _{12}$ borophene based
on the tight-binding model\cite{Boro}. In particular we explore the band
touching problem at high symmetry points. When we assume an identical
transfer energy $t_{ij}$ and an on-site energy $\varepsilon _{i}$, massless
Dirac fermions emerge at two-band touching points ($K$ and $K^{\prime }$\
points), and different types of fermions emerge at three-band touching
points ($X$, $M$, $\Lambda $ and $\Lambda ^{\prime }$\ points). In
particular, fermions at the $X$ and $M$ points constitute
pseudospin triplets separately. Then we construct an effective two-band
theory or three-band theory in the vicinity of each touching point. Next, we
consider the models together with realistic parameters for $t_{ij}$ and $%
\varepsilon _{i}$. We consider two models with and without the inversion
symmetry by an appropriate choice of the on-site energies. We find
anisotropic massive Dirac fermions with the use of inversion nonsymmetric
parameters. The degeneracy at the three-band touching points are slightly
resolved both for the inversion symmetric and nonsymmetric models. Finally
we study the edge states of borophene nanoribbons.

This paper is composed as follows. In Sec. II we review the basic properties
of the lattice structure, the Brillouin zone and the symmetry for $\beta
_{12}$ borophene. In Sec. III, we start with a five-band model comprised of
the $p_{z}$ orbitals of Boron. We compare three types of models. One is a
homogeneous model, where we take an identical transfer energy and an
identical on-site energy. The second is the inversion symmetric model, where
the transfer and on-site energies are chosen so as to respect the inversion
symmetry. The third is the inversion nonsymmetric model, where on-site
energies breaks the inversion symmetry. We show that Dirac fermions are
gapless (gapped) when the inversion symmetry is present (absent). In Sec.IV,
we derive an effective Dirac theory for general parameters and confirm the
above results. In Sec.V, we derive effective theories of fermions at
three-band touching points. It is shown that the set of fermions at
the $X$ or $M$ point is unitary equivalent to the triplet of the pseudospin (%
$J=\pm 1,0$). In Se.VI, we study edge states of borophene nanoribbons, where
five different types of edges are introduced corresponding to the unit cell
number.

\begin{figure}[t]
\centerline{\includegraphics[width=0.5\textwidth]{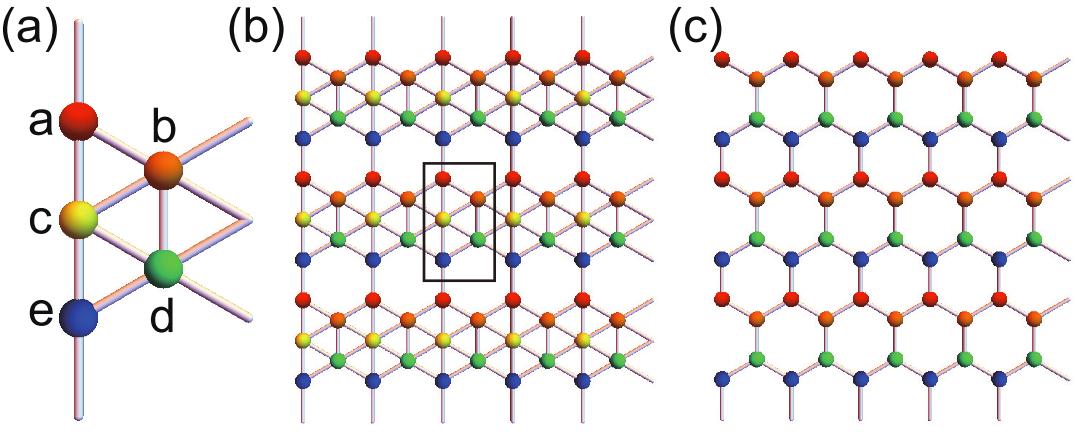}}
\caption{(a) The unit cell of borophene, which contains five atoms. Each
atoms are colored by red, orange, yellow, green and blue, which we label
"a", "b", "c", "d" and "e" atoms, respectively. (b) The lattice structure of
borophene. The unit cell is indecated by the black rectangle. (c) The
honeycomb lattice obtained by removing the "c" atoms, which describes the
effective four-band and two-band model. }
\label{FigLattice}
\end{figure}

\section{$\protect\beta_{12}$ lattice}

The lattice structure of $\beta _{12}$ borophene is illustrated in Fig.\ref%
{FigLattice}(b). The unit cell contains five atoms as in Fig.\ref{FigLattice}%
(a). The "a" and "e" atoms have four bonds, the "b" and "d" atoms have five
bonds, and the "c" atoms have six bonds, leading to different on-site
potentials.

The Brillouin zone is a rectangular given by $-\pi /a\leq k_{x}\leq \pi /a$
and $-\pi /(\sqrt{3}a)\leq k_{y}\leq \pi /(\sqrt{3}a)$, as shown in Fig.\ref%
{FigBrillouin}(a). However, it is convenient to use a shifted Brillouin zone 
$0\leq k_{x}\leq 2\pi /a$ and $-\pi /(\sqrt{3}a)\leq k_{y}\leq \pi /(\sqrt{3}%
a)$. The area of the Brillouin zone of $\beta _{12}$ borophene is one half
of that of the honeycomb lattice. This is understood as follows. The $\beta
_{12}$ lattice without the "c" atoms is identical to the honeycomb lattice,
where the unit cell contains four atoms. On the other hand, the honeycomb
lattice has only two atoms in the unit cell. Namely, the Brillouin zone of
the $\beta _{12}$ borophene must be one half of that of the honeycomb
lattice.

The symmetries of the lattice are the inversion symmetry $I$, the two mirror
symmetries with respect to the $x$ and $y$ axes, $M_{x}$ and $M_{y}$. There
is a relation $I=M_{x}M_{y}$. On the other hand, the $C_{3}$ rotation
symmetry is absent in the $\beta _{12}$ lattice, which exists in the
honeycomb lattice.

\begin{figure}[t]
\centerline{\includegraphics[width=0.5\textwidth]{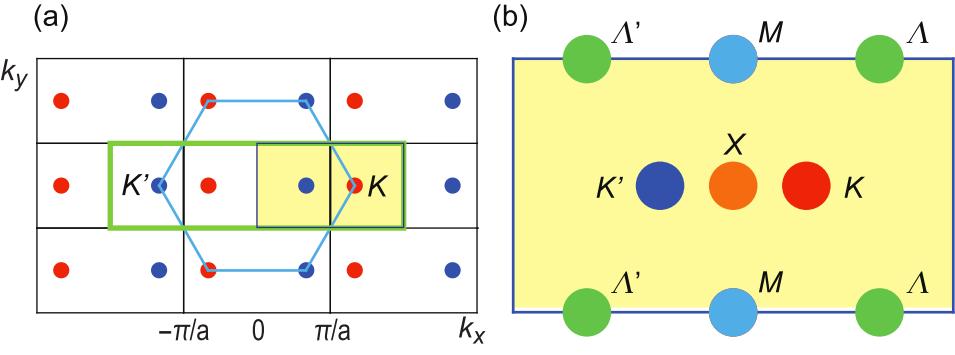}}
\caption{ (a) Brillouin zone of borophene indicated by the yellow rectangle
with the blue boundary, and that of the honeycomb lattice indicated by a
hexagon with the cyan boundary or equivalently a rectangle with the green
boundary. The area of the former is one half of the latter. (b) Positions at
which Dirac fermions ($K,K^{\prime }$) and triplet fermions ($X,M$) emerge
in the Brillouin zone. Additionally there appear three-band touching points
at $\Lambda $ and $\Lambda ^{\prime }$.}
\label{FigBrillouin}
\end{figure}

\section{Five-band model}

A recent first-principles calculation demonstrates that the system is well
described only in terms of the $p_{z}$ orbitals of the boron atoms\cite{Boro}%
, implying that the tight-binding model is five dimensional, 
\begin{equation}
H_{5}=\left( 
\begin{array}{ccccc}
\varepsilon _{a} & t_{ab}g & t_{ac}f^{\ast } & 0 & t_{ae}f \\ 
t_{ab}g^{\ast } & \varepsilon _{b} & t_{bc}g & t_{bd}f^{\ast } & 0 \\ 
t_{ac}f & t_{bc}g^{\ast } & \varepsilon _{c} & t_{cd}g & t_{ce}f \\ 
0 & t_{bd}f & t_{cd}g^{\ast } & \varepsilon _{d} & t_{de}g \\ 
t_{ae}f^{\ast } & 0 & t_{ce}f & t_{de}g^{\ast } & \varepsilon _{e}%
\end{array}%
\right)  \label{H5}
\end{equation}%
with 
\begin{equation}
f=e^{iak_{y}/\sqrt{3}},\quad g=2e^{-iak_{y}/2\sqrt{3}}\cos ak_{x}/2.
\end{equation}%
The parameters obtained by fitting first-principles calculation results are
summarized as\cite{Boro}, 
\begin{align}
t_{ab}& =t_{de}=-2.04,\quad t_{ac}=t_{ce}=-1.79,\quad t_{ae}=-2.12,  \notag
\\
t_{bc}& =t_{cd}=-1.84,\quad t_{bd}=-1.91,  \label{para}
\end{align}%
showing that the transfer energies are symmetric along the $x$ axis, and%
\begin{equation}
\varepsilon _{a}=\varepsilon _{d}=0.196,\quad \varepsilon _{b}=\varepsilon
_{e}=-0.058,\quad \varepsilon _{c}=-0.845.  \label{OnSiteBoro}
\end{equation}%
The lattice constant is $a=2.9236$\AA . A characteristic feature is that the
inversion symmetry of the lattice structure is broken by this set of on-site
energies. We refer to this tight-binding model as the inversion nonsymmetric
model.

\begin{figure}[t]
\centerline{\includegraphics[width=0.5\textwidth]{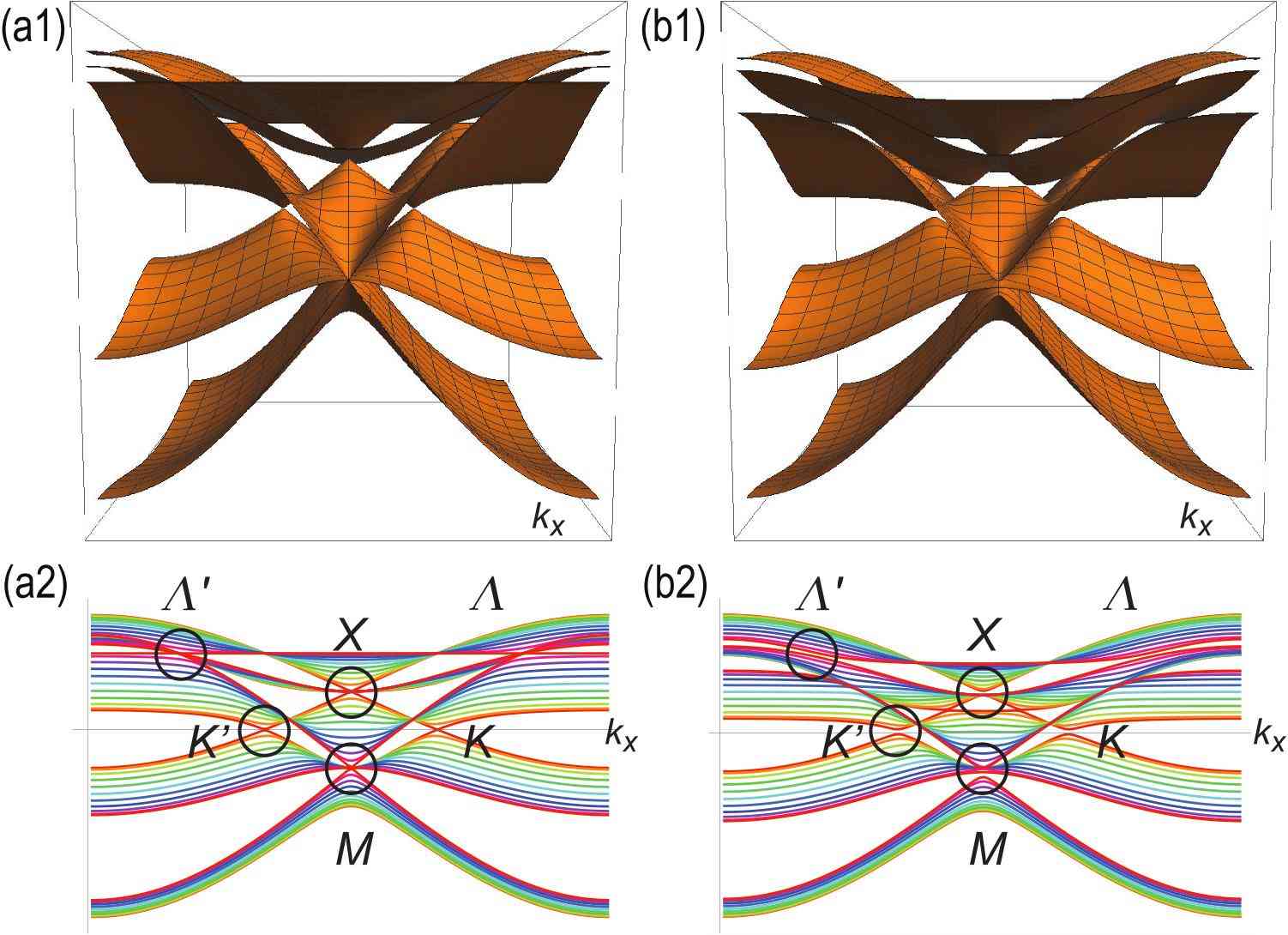}}
\caption{ Bird's eye's views of the band structure of (a1) the homogeneous
model and (b1) the inversion nonsymmetric model. The projected band
structure along the $k_{x}$ direction of (a2) the homogeneous model and (b2)
the inversion nonsymmetric model. Gapless Dirac fermions ($K,K^{\prime }$)
and gapless triplet fermions ($X,M$) emerge in the homogeneous model. In the
inversion nonsymmetric model, Dirac fermions have a tiny gap, while the
degeneracy of triplet fermions is slightly resolved. }
\label{FigProject}
\end{figure}

We first consider the model\cite{Boro} by setting all transfer energy equal (%
$t_{ij}=t=-2$eV) and all on-site energy zero ($\varepsilon _{i}=0$), which
we refer to as the homogeneous model. We also investigate the inversion
symmetric model, which is defined by the following set of the one-site
energies instead of (\ref{OnSiteBoro}), 
\begin{equation}
\varepsilon _{a}=\varepsilon _{e}=0.196,\quad \varepsilon _{b}=\varepsilon
_{d}=-0.058,\quad \varepsilon _{c}=-0.845,  \label{OnSiteSym}
\end{equation}%
where the magnitude of the on-site energy reflects the number of adjacent
atoms in each sites. We show the band structures of the homogeneous model
and the inversion nonsymmetric model in Fig.\ref{FigProject}(a1) and Fig.\ref%
{FigProject}(b1), respectively. We also show their project band structures
along the $k_{x}$ axis in Fig.\ref{FigProject}(a2) and Fig.\ref{FigProject}%
(b2). Those for the inversion symmetric model are quite similar to these.

\begin{figure}[t]
\centerline{\includegraphics[width=0.5\textwidth]{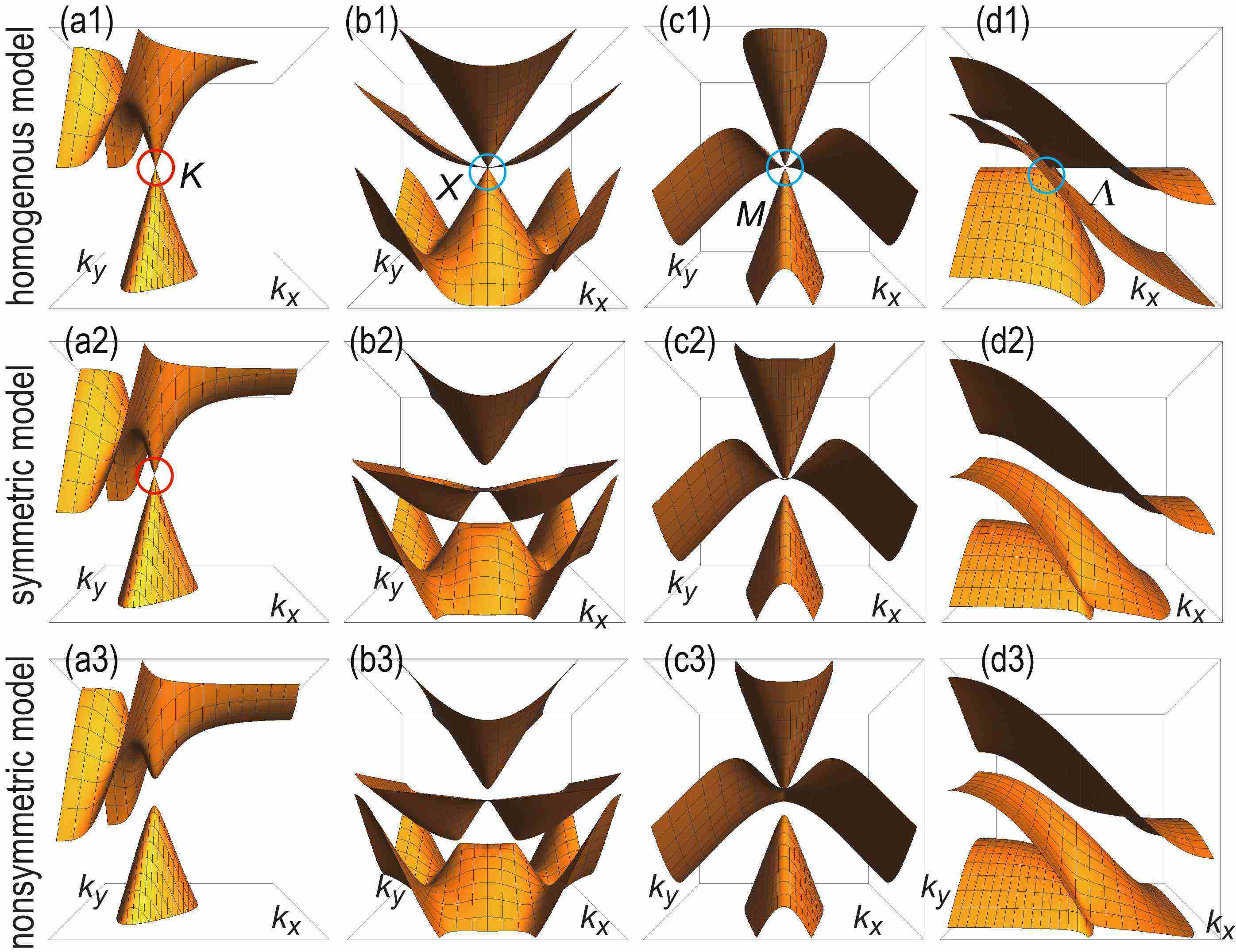}}
\caption{ Band structures in the vicinity of the Dirac point ($K$) indicated
by a magenta circle and the three-band touching points ($X,M,\Lambda $)
indicated by cyan circles: (a1)$\sim $(d1) for the homogeneous model; (a2)$%
\sim $(d2) for the inversion symmetric model; (a3)$\sim $(d3) for the
inversion nonsymmetric model. Dirac fermions remains gapless while the
three-band touching are slightly resolved in the inversion symmetric model.
Dirac fermions become gapped in the inversion nonsymmetric model.}
\label{FigZoom}
\end{figure}

We start with the investigation of the homogeneous model: See Fig.\ref%
{FigProject}(a1). We find Dirac fermions at the two points $\mathbf{K}_{\pm
}=\left( k_{x},k_{y}\right) =\left( \pm 2\pi /\left( 3a\right) ,0\right) $,
where the energy is explicitly obtained as 
\begin{equation}
U_{1}^{-1}H_{5}\left( \mathbf{K}_{\pm }\right) U_{1}=t\,\text{diag.}\left(
0,0,1\pm \sqrt{5},-2\right)
\end{equation}%
with the use of a unitary transformation $U_{1}$. (These two points are
customarily called the $K$ and $K^{\prime }$ points.) It is interesting that
there are different types of three-band touching points. Their positions in
the Brillouin zone are shown in Fig.\ref{FigBrillouin}(b). One is at the
point $\mathbf{X}=\left( \pi /a,0\right) $, where the energy is given by 
\begin{equation}
U_{2}^{-1}H_{5}\left( \mathbf{X}\right) U_{2}=t\,\text{diag.}\left(
-1,-1,-1,1,2\right) .
\end{equation}%
The second one is at the point $\mathbf{M}=\left( \pi /a,\pi /\sqrt{3}%
a\right) $, where the energy is given by 
\begin{equation}
U_{3}^{-1}H_{5}\left( \mathbf{M}\right) U_{3}=t\,\text{diag.}\left(
1,1,1,-1,-2\right) .
\end{equation}%
The third one is at the points $\mathbf{\Lambda }_{\pm }=\left( \pm \pi
/\left( 3a\right) ,\pi /\left( \sqrt{3}a\right) \right) $, where the energy
is given by 
\begin{equation}
U_{4}^{-1}H_{5}\left( \mathbf{\Lambda }_{\pm }\right) U_{4}=t\,\text{diag.}%
\left( -2,-2,-2,2,4\right) .
\end{equation}%
We show the detailed band structures of Dirac fermions and triple-point
fermions in Fig.\ref{FigZoom}(a1),(b1),(c1) and (d1).

We may investigate both the inversion symmetric and nonsymmetric models in a
similar way. Their overall band structures are very similar to that of the
homogeneous model, as shown in Fig.\ref{FigProject}(b1) for the inversion
nonsymmetric model. However, there arise differences with respect to the
degeneracy at the band touching points. First, the Dirac fermions remains
gapless in the inversion symmetric model but gets gapped in the inversion
nonsymmetric model. On the other hand, three-band touching points are
slightly resolved both in the inversion symmetric and nonsymmetric models.
We show the detailed band structure at these points in Fig.\ref{FigZoom}.

\section{Dirac fermions}

Since the dimension of the matrix in the tight-binding model (\ref{H5}) is
five, it is impossible to diagonalize it analytically. It is highly
desirable to construct such models with lower dimensions that we can analyze
analytically. We construct an effective two-band Hamiltonian for Dirac
fermions in this section and effective three-band Hamiltonians for 
fermions at three-band touching points in the next section.

\begin{figure}[t]
\centerline{\includegraphics[width=0.5\textwidth]{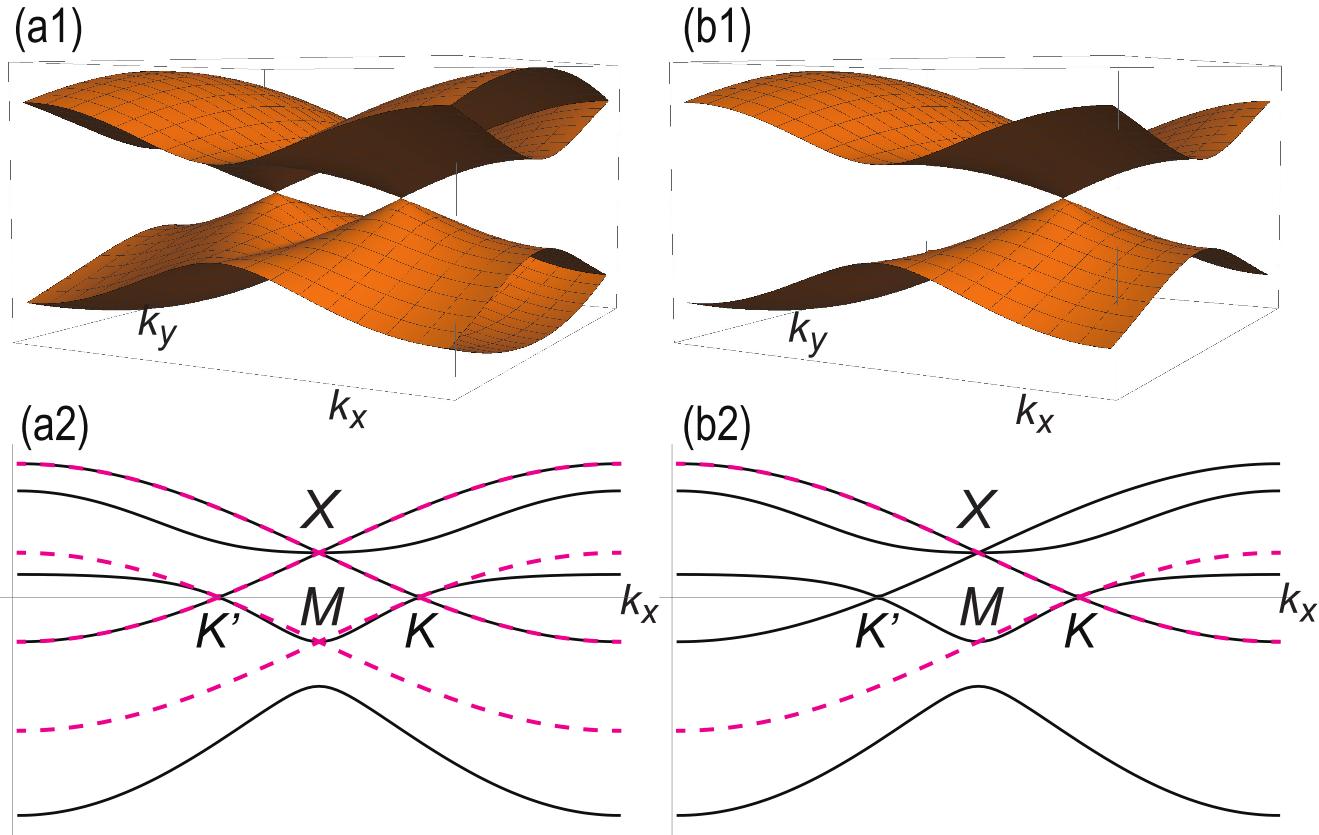}}
\caption{ Bird's eye's views of the band structure of (a1) the effective
four-band model and (b1) the effective two-band model. The band structure
indicated in magenta along the $k_{x}$ direction at $k_{y}=0$ for (a2) the
effective four-band model and (b2) the effective two-band model together
with the band structure of the five-band model indicated in black. }
\label{Fig4Band}
\end{figure}

It is reported\cite{Boro} that the amplitude of the zero-energy wave
function at the $K$ and $K^{\prime }$ points is exactly zero at the "c"
sites for the homogeneous model. Then, it is reasonable to neglect the "c"
atoms in the Hamiltonian (\ref{H5}), and we obtain the following four-band
Hamiltonian to describe the physics in the vicinity of the $K$ and $%
K^{\prime }$ points, 
\begin{equation}
H_{5}=\left( 
\begin{array}{cccc}
0 & g & 0 & f \\ 
g^{\ast } & 0 & f^{\ast } & 0 \\ 
0 & f & 0 & g \\ 
f^{\ast } & 0 & g^{\ast } & 0%
\end{array}%
\right) .
\end{equation}%
The energy is analytically obtained as%
\begin{align}
E& =\pm \sqrt{f-g}\sqrt{f^{\ast }-g^{\ast }},\pm \sqrt{f+g}\sqrt{f^{\ast
}+g^{\ast }}  \notag \\
& =\pm t\sqrt{3+2\cos ak_{x}\pm 2\sqrt{\left( 1+\cos ak_{x}\right) \left(
1+\cos \sqrt{3}ak_{y}\right) }}.
\end{align}%
Indeed, as we show in Fig.\ref{Fig4Band}(a), it well reproduces the original
band structure of the Dirac fermions both at the $K$ and $K^{\prime }$
points. 

The corresponding lattice is a honeycomb lattice shown in Fig.\ref%
{FigLattice}(c). The unit cell contains four atoms comprised of the "a",
"b", "d" and "e" atoms. On the other hand, the unit cell of the honeycomb
lattice contains two atoms such as "b" and "d". Hence the above four-band
model can be further reduced to the two-band model.

We are able to construct actually the two-band model by way of $H_{2}(%
\mathbf{k})=P_{2}U_{1}^{-1}H_{5}(\mathbf{k})U_{1}P_{2}$ in the vicinity of
the Dirac points, where $P_{2}$ is the projection operator from the $5\times
5$ Hamiltonian to the $2\times 2$ Hamiltonian containing the two bands with
the zero eigen-energy. The low-energy effective Hamiltonian is given by%
\begin{equation}
H_{2}=\left( 
\begin{array}{cc}
\left( \varepsilon _{b}+\varepsilon _{e}\right) /2 & F \\ 
F^{\ast } & \left( \varepsilon _{a}+\varepsilon _{d}\right) /2%
\end{array}%
\right)  \label{H2}
\end{equation}%
with%
\begin{equation}
F=\frac{t_{ae}+t_{bd}}{2}e^{-iak_{y}/\sqrt{3}}+\left( t_{ab}+t_{de}\right)
e^{iak_{y}/2\sqrt{3}}\cos \frac{ak_{x}}{2}.
\end{equation}%
It is identical to the Hamiltonian of the anisotropic honeycomb lattice with
on-site potentials,\cite{Phos} which corresponds to the lattice without the
"c" atoms. This correspondence is due to the fact that the amplitude of the
wave function at the "c" sites is zero at the zero energy.

The gap closes at 
\begin{equation}
\tilde{K}=\left( \frac{2}{a}\arccos \frac{-\left( t_{ae}+t_{bd}\right) }{%
2\left( t_{ab}+t_{de}\right) },0\right) ,
\end{equation}%
when $\varepsilon _{a}-\varepsilon _{b}+\varepsilon _{d}-\varepsilon _{e}=0$%
. Especially, the gap closes in the presence of the inversion symmetry since 
$\varepsilon _{a}=\varepsilon _{e}$ and $\varepsilon _{b}=\varepsilon _{d}$.
The gap closing point shifts from the original $K$ point when $%
t_{ae}+t_{bd}\neq t_{ab}+t_{de}$. The gap opens when $\varepsilon
_{a}-\varepsilon _{b}+\varepsilon _{d}-\varepsilon _{e}\neq 0$ with the gap $%
\left\vert \varepsilon _{a}-\varepsilon _{b}+\varepsilon _{d}-\varepsilon
_{e}\right\vert /2$. We estimate the gap is $0.254$eV by using the
parameters (\ref{OnSiteBoro}) in the inversion nonsymmetric model.

We show the band structure along the $k_{x}$ axis of the two-band model as
well as the five-band model in Fig.\ref{Fig4Band}(b). The Dirac fermions at
the $K$ point is well reproduced, while the Dirac fermions at the $K^{\prime
}$ point disappear. This is due to the fact that the Brillouin zone is
enlarged twice since the unit cell becomes half compared with that of the
four-band model, as shown in Fig.\ref{FigBrillouin}. It is necessary to
construct another two-band model by expanding at the $K^{\prime }$ point,
which is done precisely in the similar way.

In the vicinity of the $K$ point, the two-band Hamiltonian is expanded as%
\begin{align}
H_{2}^{K}& =-\frac{\varepsilon _{a}-\varepsilon _{b}+\varepsilon
_{d}-\varepsilon _{e}}{4}\tau _{z}-\frac{\varepsilon _{a}-\varepsilon
_{b}+\varepsilon _{d}-\varepsilon _{e}}{4}\tau _{0}  \notag \\
& -\frac{\sqrt{4\left( t_{ab}+t_{de}\right) ^{2}-\left( t_{ae}+t_{bd}\right)
^{2}}}{4}a\left( k_{x}-\tilde{K}\right) \tau _{x}  \notag \\
& +\frac{\sqrt{3}}{4}\left( t_{ae}+t_{bd}\right) ak_{y}\tau _{y},
\end{align}%
which describe a Dirac cone. The dispersion is isotropic only for the
homogeneous model.

In passing, it is intriguing to see that the four-band theory describes
precisely two bands of the $X$ and $M$ points though it is constructed soley
for the $K$ and $K^{\prime }$ points in Fig.\ref{Fig4Band}.

\section{Fermions at three-band touching points}

We next construct three-band effective models for fermions at
three-band touching points. We construct effective models as in the case of
the Dirac fermions with the use of the unitary transformation and the
projection to the low-energy bands.

\textit{X point:} The effective Hamiltonian valid in the vicinity of the $X$
point is given by 
\begin{equation}
H_{3}^{X}=t\left( 
\begin{array}{ccc}
F_{aa}^{X} & F_{ab}^{X} & F_{ac}^{X} \\ 
F_{ab}^{X\ast } & F_{bb}^{X} & F_{bc}^{X} \\ 
F_{ac}^{X\ast } & F_{bc}^{X\ast } & F_{cc}^{X}%
\end{array}%
\right) ,  \label{HA}
\end{equation}%
where 
\begin{align}
F_{aa}^{X}=& F_{bb}^{X}=F_{cc}^{X}=-\cos \frac{ak_{y}}{\sqrt{3}},  \notag \\
F_{ab}^{X}=& 2\cos \frac{ak_{x}}{2}\cos \frac{ak_{y}}{2\sqrt{3}},\quad
F_{ac}^{A}=i\sqrt{3}\sin \frac{ak_{y}}{\sqrt{3}},  \notag \\
F_{bc}^{X}=& 2i\sqrt{3}\cos \frac{ak_{x}}{2}\sin \frac{ak_{y}}{2\sqrt{3}}
\end{align}%
in the case of the homogeneous model. See Appendix for general parameters.

\begin{figure}[t]
\centerline{\includegraphics[width=0.5\textwidth]{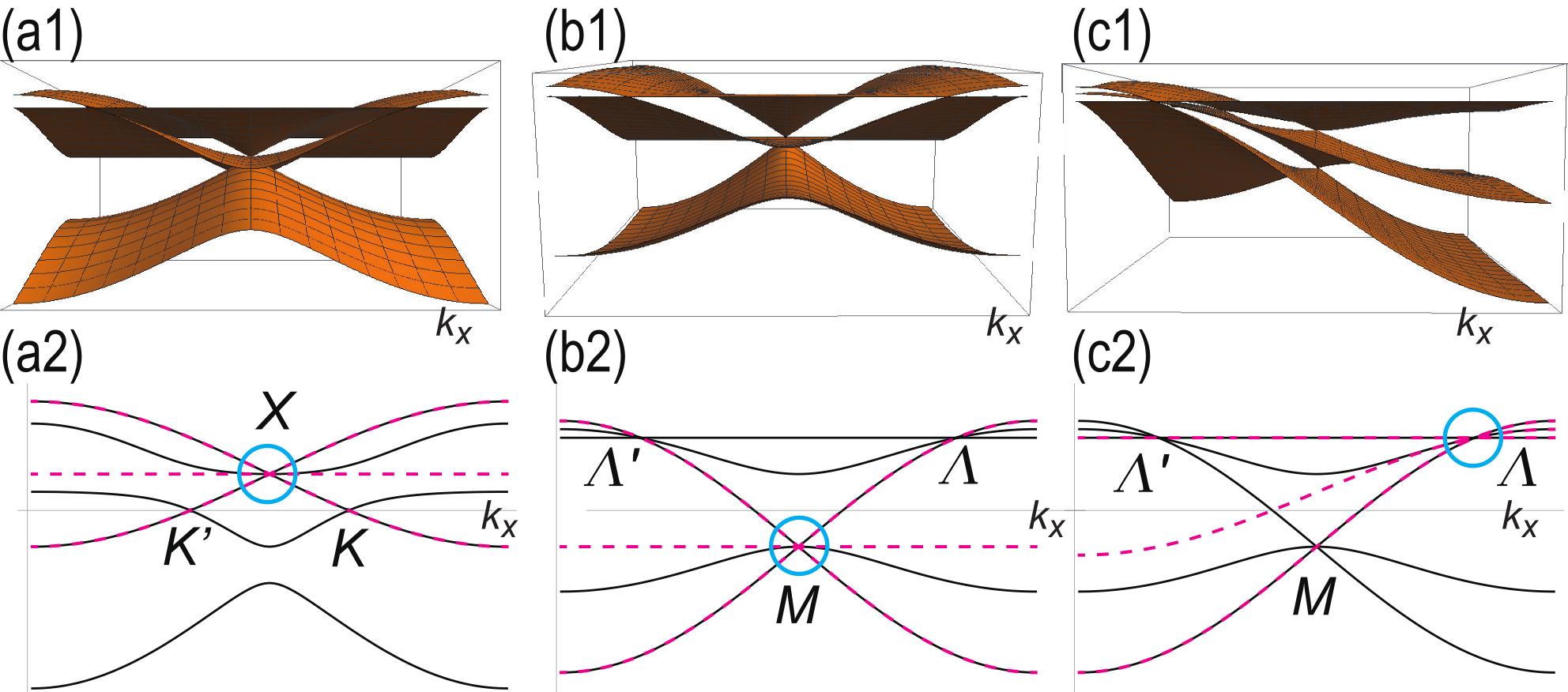}}
\caption{ Bird's eye's views of the band structure of the effective
three-band models describing fermions at (a1) the $X$ point, (b1) the $M$
point and (c1) the $\Lambda $ point. (a2)$\sim$(c2) Black curves represent
the band structure of the five-band model (\protect\ref{H5}) along the $k_{x}
$ axis. Magenta curves represent that of (a2) the two-band model (\protect
\ref{H2}) at the $X$ point ($k_{y}=0$), (b2) the three-band model (\protect
\ref{HA}) at the $M$ point ($k_{y}=\protect\pi /(\protect\sqrt{3}a)$) and
(c2) the three-band model (\protect\ref{HB}) at the $\Lambda $ point ($k_{y}=%
\protect\pi /(\protect\sqrt{3}a)$). }
\label{FigEfBand}
\end{figure}

We show the band structure in Fig.\ref{FigEfBand}(a1). The vicinity of the
three-band touching point is well reproduced by this model. Furthermore,
comparing the band structure of the three-band model $H_{3}^{X}$\ along the $%
k_{y}=0$\ line with that of the original five-band model $H_{5}$, the
three-band model $H_{3}^{X}$ is found to reproduce perfectly the two bands
given by $E=\pm 2\cos (ak_{x}/2)-1$ all over the region: See \ref{FigEfBand}%
(a2).

We wonder why there is no partner for the $X$\ point as in the case of the $%
K $\ and $K^{\prime }$\ points. We study this problem for the homogeneous
model. The three-band Hamiltonian is expanded in the vicinity of the $X$\
point as%
\begin{equation}
H_{3}^{X}=-t+ta\left( 
\begin{array}{ccc}
0 & -k_{x}^{\prime } & ik_{y} \\ 
-k_{x}^{\prime } & 0 & 0 \\ 
-ik_{y} & 0 & 0%
\end{array}%
\right)  \label{HamilX}
\end{equation}%
with $k_{x}^{\prime }=k_{x}\mp \frac{\pi }{a}$. The corresponding wave
functions are%
\begin{align}
\psi _{0}& =\left\{ 0,i\sin \theta ,\cos \theta \right\} ^{t}, \\
\psi _{\pm }& =\frac{1}{\sqrt{2}}\left\{ \pm i,-i\cos \theta ,\sin \theta
\right\} ^{t}.
\end{align}%
The Berry phase is zero for each band,%
\begin{equation}
\Gamma _{\text{B}}=-i\int d\theta \left\langle \psi _{0}\right\vert \frac{%
\partial }{\partial \theta }\left\vert \psi _{0}\right\rangle =-i\int
d\theta \left\langle \psi _{\pm }\right\vert \frac{\partial }{\partial
\theta }\left\vert \psi _{\pm }\right\rangle =0.
\end{equation}%
Since the band carries no topological charge, the $X$ point can exist by
itself. Furthermore, it indicates that the three-band touching point is not
topologically protected.

The energy eigenstate of the Hamiltonian (\ref{HamilX}) is given by $%
E=-t,-t\pm \sqrt{k_{x}^{\prime 2}+k_{y}^{2}}$. Hence it is unitary
equivalent to the following Hamiltonian,%
\begin{equation}
H=-t+ta(k_{x}^{\prime }J_{x}+k_{y}J_{y}),
\end{equation}%
where $\mathbf{J}=(J_{x},J_{y},J_{z})$ is the pseudospin operator obeying $%
[J_{x},J_{y}]=J_{z}$, etc., whose magnitude is $J=1$. Namely, the three
bands are members of a pseudospin triplet with $J=\pm1,0$. On the other
hand, for realistic $t_{ij}$ and $\varepsilon _{i}$, the three-band touching
point is resolved.

\textit{M point:} In the vicinity of the $M$ point, we obtain the following
effective three-band model, 
\begin{equation}
H_{3}^{M}=t\left( 
\begin{array}{ccc}
F_{aa}^{M} & F_{ab}^{M} & F_{ac}^{M} \\ 
F_{ab}^{M\ast } & F_{bb}^{M} & F_{bc}^{M} \\ 
F_{ac}^{M\ast } & F_{bc}^{M\ast } & F_{cc}^{M}%
\end{array}%
\right) ,  \label{HB}
\end{equation}%
where 
\begin{align}
F_{aa}^{M}=& F_{bb}^{M}=F_{cc}^{M}=\frac{1}{2}\left( \cos \frac{ak_{y}}{%
\sqrt{3}}+\sqrt{3}\sin \frac{ak_{y}}{\sqrt{3}}\right) ,  \notag \\
F_{ab}^{M}=& \left( -1\right) ^{2/3}\cos \frac{ak_{x}}{2}\left( \sqrt{3}\sin 
\frac{ak_{y}}{2\sqrt{3}}-\cos \frac{ak_{y}}{2\sqrt{3}}\right) ,  \notag \\
F_{ac}^{M}=& \frac{\sqrt{3}}{2}\left( -1\right) ^{5/6}\left( \sin \frac{%
ak_{y}}{\sqrt{3}}-\sqrt{3}\cos \frac{ak_{y}}{\sqrt{3}}\right) ,  \notag \\
F_{bc}^{M}=& \left( -1\right) ^{1/6}\sqrt{3}\cos \frac{ak_{x}}{2}\left( \sin 
\frac{ak_{y}}{2\sqrt{3}}+\sqrt{3}\cos \frac{ak_{y}}{2\sqrt{3}}\right)
\end{align}%
in the case of the homogeneous model. See Appendix for general parameters.

We show the band structure in Fig.\ref{FigEfBand}(b1). In Fig.\ref{FigEfBand}%
(b2), comparing the band structure of the three-band model $H_{3}^{M}$ along
the $k_{y}=\pi /\sqrt{3}a$ line with that of the five-band model $H_{5}$, we
find that the three-band model $H_{3}^{M}$ perfectly reproduces the
two-bands given by $E=t\pm 2\cos \frac{ak_{x}}{2}$ all over the region. On
the other hand, the middle band becomes a perfect flat band in the
three-band model $H_{3}^{M}$, while it is dispersive in the five-band model $%
H_{5}$.

We expand the three-band Hamiltonian in the vicinity of the $\mathbf{M}$
point as 
\begin{equation}
H_{3}^{M}=t+ta\left( 
\begin{array}{ccc}
0 & 0 & e^{5i\pi /6}k_{y}^{\prime } \\ 
0 & 0 & -\sqrt{3}e^{i\pi /6}k_{x}^{\prime } \\ 
e^{-5i\pi /6}k_{y}^{\prime } & -\sqrt{3}e^{-i\pi /6}k_{x}^{\prime } & 0%
\end{array}%
\right)
\end{equation}%
with $k_{x}^{\prime }=k_{x}-\pi /a,\quad k_{y}^{\prime }=k_{y}-\pi /\left( 
\sqrt{3}a\right) $. The energy is obtained as $E=t,t\pm \sqrt{3k_{x}^{\prime
2}+k_{y}^{\prime 2}}$. Hence it is unitary equivalent to the following
Hamiltonian,%
\begin{equation}
H=t+ta(\sqrt{3}k_{x}^{\prime }J_{x}+k_{y}J_{y}).
\end{equation}%
The three bands are members of a pseudospin triplet with $J=\pm 1,0$. The
Berry phase is zero for each band. Consequently, the $M$ point is not
accompanied by a partner. These properties are quite similar to those of the 
$X$ point.

\begin{figure}[t]
\centerline{\includegraphics[width=0.5\textwidth]{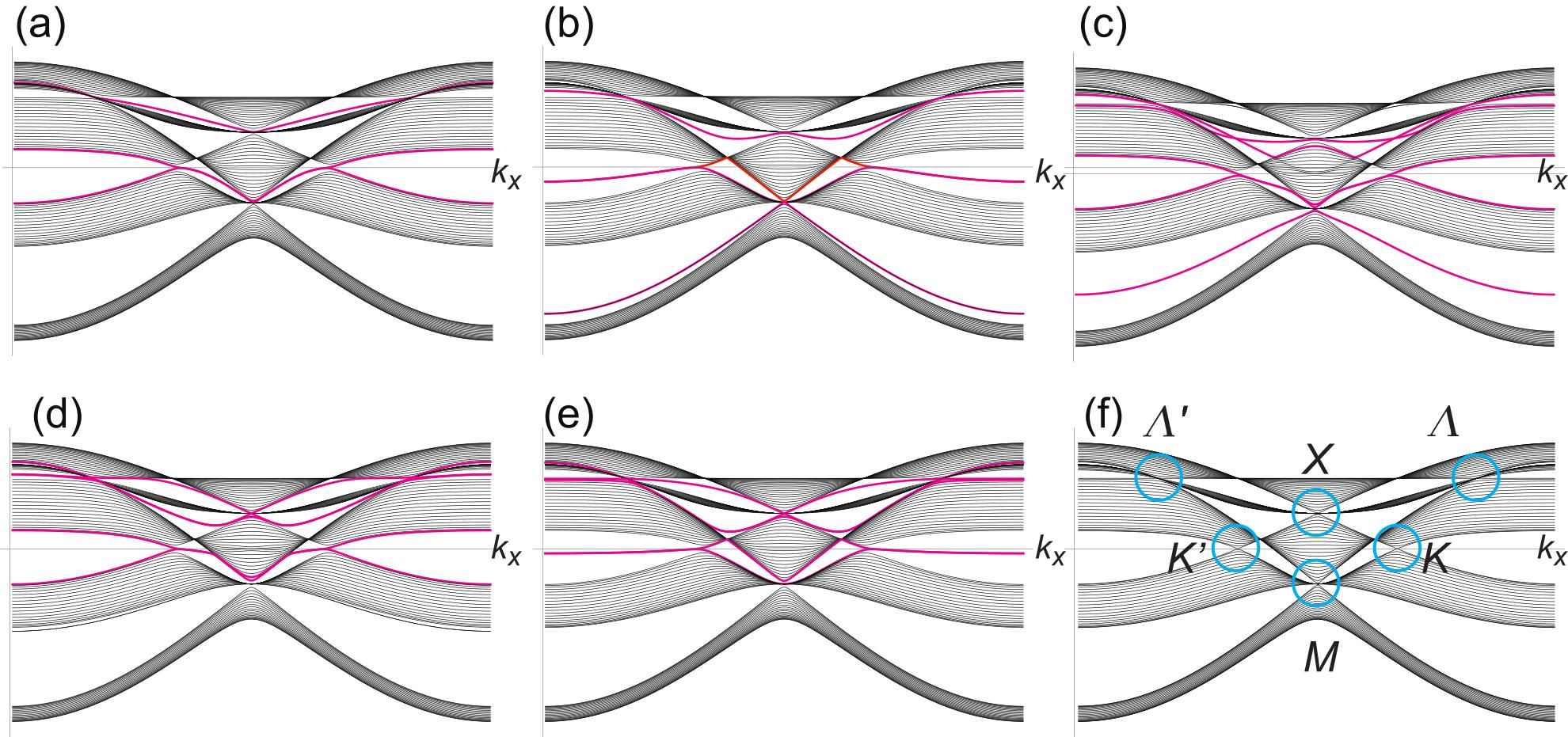}}
\caption{ Band structures of zigzag borophene nanoribbon. (a) The upper edge
is terminated by "a" atoms, while the lowe edge is terminated by "e" atoms.
(b) The upper edge is terminated by "b" atoms, while the lowe edge is
terminated by "d" atoms. (c) Both edges are terminated by "c" atoms. (d) The
upper edge is terminated by "d" atoms, while the lowe edge is terminated by
"b" atoms. (e) The upper edge is terminated by "e" atoms, while the lowe
edge is terminated by "a" atoms. One edge state connects a Dirac point and a
triple point. The edge states are colored in magenta. (f) The projected band
structure of the bulk along the $k_{x}$ axis, which is the same one as Fig.%
\protect\ref{FigProject}(a2). }
\label{FigRibbon}
\end{figure}

$\Lambda $ point\textit{:} In the vicinity of the $\Lambda $ point, we
obtain the following effective three-band model, 
\begin{equation}
H_{3}^{\Lambda }=t\left( 
\begin{array}{ccc}
F_{aa}^{\Lambda } & F_{ab}^{\Lambda } & F_{ac}^{\Lambda } \\ 
F_{ab}^{\Lambda \ast } & F_{bb}^{\Lambda } & F_{bc}^{\Lambda } \\ 
F_{ac}^{\Lambda \ast } & F_{bc}^{\Lambda \ast } & F_{cc}^{\Lambda }%
\end{array}%
\right)
\end{equation}%
with%
\begin{align}
F_{aa}^{\Lambda }=& F_{bb}^{\Lambda }=F_{cc}^{\Lambda }=-\frac{1}{4}[\cos 
\frac{ak_{y}}{\sqrt{3}}+\sqrt{3}\sin \frac{ak_{y}}{\sqrt{3}}  \notag \\
& +2\sqrt{3}\cos \frac{ak_{x}}{2}\left( \sqrt{3}\cos \frac{ak_{y}}{2\sqrt{3}}%
+\sin \frac{ak_{y}}{2\sqrt{3}}\right) ],  \notag \\
F_{ab}^{\Lambda }=& \frac{1}{4}\left( -1\right) ^{2/3}[-3\cos \frac{ak_{y}}{%
\sqrt{3}}+\sqrt{3}\sin \frac{ak_{y}}{\sqrt{3}}  \notag \\
& +2\cos \frac{ak_{x}}{2}\left( \sqrt{3}\sin \frac{ak_{y}}{2\sqrt{3}}-\cos 
\frac{ak_{y}}{2\sqrt{3}}\right) ],  \notag \\
F_{ac}^{\Lambda }=& \frac{\sqrt{3}}{2}e^{-iak_{y}/\sqrt{3}}\left( -1\right)
^{2/3}+\frac{\sqrt{3}}{4}e^{iak_{y}/\sqrt{3}}  \notag \\
& -\frac{3i}{2}e^{-iak_{y}/2\sqrt{3}}\cos \frac{ak_{x}}{2},  \notag \\
F_{bc}^{\Lambda }=& \left[ \frac{\sqrt{3}}{2}\left( -1\right)
^{1/6}e^{-iak_{y}/2\sqrt{3}}+\sqrt{3}e^{iak_{y}/2\sqrt{3}}\right] \cos \frac{%
ak_{x}}{2}  \notag \\
& +\frac{3}{4}\left( -1\right) ^{1/6}e^{iak_{y}/\sqrt{3}}.
\end{align}

We show the band structure in Fig.\ref{FigEfBand}(c1). In Fig.\ref{FigEfBand}%
(c2), we compare the band structure of the three-band model $H_{3}^{C}$
along the $k_{y}=\pi /\sqrt{3}a$ line with that of the original five-band
model $H_{5}$, and find that the three band model $H_{3}^{C}$ perfectly
reproduces the two bands given by $E=-2t,t-2\cos \frac{ak_{x}}{2}$ all over
the region.

In the vicinity of the $\Lambda $ point, the Hamiltonian is expanded as%
\begin{equation}
H_{3}^{\Lambda }=-2t+\frac{\sqrt{3}}{4}ak_{x}+\frac{ta}{4}\left( 
\begin{array}{ccc}
0 & F_{ab}^{C^{\prime }} & F_{ac}^{C^{\prime }} \\ 
F_{ab}^{C^{\prime }\ast } & 0 & F_{bc}^{C^{\prime }} \\ 
F_{ac}^{C^{\prime }\ast } & F_{bc}^{C^{\prime }\ast } & 0%
\end{array}%
\right)
\end{equation}%
with%
\begin{align}
F_{ab}^{C^{\prime }} =&\left( -1\right) ^{2/3}\sqrt{3}k_{y}  \notag \\
F_{ac}^{C^{\prime }} =&\frac{1}{2}\left( -1\right) ^{1/3}\left(
3k_{x}+ik_{y}\right)  \notag \\
F_{bc}^{C^{\prime }} =&-\frac{\sqrt{3}}{2}\left( -1\right) ^{1/6}k_{-}
\end{align}%
and $k_{x}^{\prime }=k_{x}-\pi /(3a),\quad k_{y}^{\prime }=k_{y}-\pi /\left( 
\sqrt{3}a\right) $.

\section{Borophene nanoribbons}

When a nanoribbons is along the zigzag direction there are zigzag and beard
edges in the case of the honeycomb system. This is because there are two
atoms in the unit cell. There are five types of borophene nanoribbons with
zigzag edges corresponding to the fact that the unit cell contains five
atoms. For example, the edge terminated by "a","b", "c" and "d" atoms forms
a zigzag edge, while that terminated by "e" atoms forms a beard edge. Thus
there are $5\times 5=25$ different nanoribbons.

We show the band structure of typical nanoribbons in Fig.\ref{FigRibbon}(a)$%
\sim$(e). In Fig.\ref{FigRibbon}(f), we show the bulk band structure
projected to the $k_{x}$ axes for the sake of comparison. The band structure
of nanoribbons are almost identical to those of the projected bulk band
structure except for the edge states, which are marked by magenta curves.
The edge states emerge in the region connecting between the Dirac point and
the triple point. Among them, there emerge almost flat bands at the zero
energy in Fig.\ref{FigRibbon}(e), which corresponds to the beard edge
states. If the two terminations are different, the edge states are the sum
of the two terminations.

\section{Discussion}

It is interesting that Dirac fermions or triplet fermions emerge in the $%
\beta _{12}$ structure of borophene at the high symmetry points.
There is a distinctive difference between them. On one hand, Dirac fermions
emerge always in a pair: They emerge at the $K$ and $K^{\prime }$ points
just as in graphene. The reason is the Nielsen-Ninomiya theorem. Namely, the
gapless Dirac fermion has $\pm \pi $ Berry phase while the gapped Dirac
fermion has $\pm 1/2$ Chern number. They appear in a pair so that the total
topological number must be zero.

One the other hand, this is not the case for triplet fermions. There
are no partners for the $X$ and $M$ points. Indeed, calculating the
Berry phases of the bands at $X$ and $M$, we find them to be zero.

The lattice structure of $\beta _{12}$ borophene has the inversion symmetry,
where massless Dirac fermions are expected. However, the inversion symmetry
is broken in the Hamiltonian together with the parameters (\ref{para})
presented in Ref.\cite{Boro}, where Dirac fermions are gapped. The ARPES
experiment shows that the gap is absent within the experimental resolution.

We note that there is a metallic band at the zero-energy in the five-band
model, while it is absent in the effective lower-band theories. It is
because that they are valid in the vicinity of the high symmetry points. We
should use the five-band model when we calculate the conductivity and others.

The author is very much grateful to N. Nagaosa for many helpful discussions
on the subject. He thanks the support by the Grants-in-Aid for Scientific
Research from MEXT KAKENHI (Grant Nos.JP25400317 and JP15H05854). This work
is also supported by JST, CREST (Grant No. JPMJCR16F1).

\appendix

\section*{Appendix: Three-band theories with general parameters}

In this appendix we present the matrix elements of the effective three-band
theories (\ref{HA}) and (\ref{Fig4Band}) with general parameters $t_{ij}$
and $\varepsilon _{i}$.

\begin{widetext}

X point:%
\begin{align}
F_{aa}^{X}& =\frac{\varepsilon _{a}+\varepsilon _{e}}{2}-t_{ae}\cos \frac{%
ak_{y}}{\sqrt{3}},  \notag \\
F_{bb}^{X}& =\frac{\varepsilon _{b}+\varepsilon _{d}}{2}-t_{bd}\cos \frac{%
ak_{y}}{\sqrt{3}},  \notag \\
F_{cc}^{X}& =\frac{\varepsilon _{a}+4\varepsilon _{c}+\varepsilon _{e}}{6}-%
\frac{2t_{ac}-t_{ae}+2t_{ce}}{3}\cos \frac{ak_{y}}{\sqrt{3}},  \notag \\
F_{ab}^{X}& =\left( t_{ab}e^{-\frac{iak_{y}}{2\sqrt{3}}}+t_{cd}e^{\frac{%
iak_{y}}{2\sqrt{3}}}\right) \cos \frac{ak_{x}}{2},  \notag \\
F_{ac}^{X}& =\frac{\varepsilon _{a}-\varepsilon _{e}}{2\sqrt{3}}+\frac{%
-t_{ac}+t_{ce}}{\sqrt{3}}\cos \frac{ak_{y}}{\sqrt{3}}+i\frac{%
t_{ac}+t_{ae}+t_{ce}}{\sqrt{3}}\sin \frac{ak_{y}}{\sqrt{3}},  \notag \\
F_{bc}^{X}& =\frac{-\left( t_{de}+2t_{bc}\right) e^{-\frac{iak_{y}}{2\sqrt{3}%
}}+\left( t_{ab}+2t_{cd}\right) e^{\frac{iak_{y}}{2\sqrt{3}}}}{\sqrt{3}}\cos 
\frac{ak_{x}}{2}.
\end{align}

M point:%
\begin{align}
F_{aa}^{M}& =\frac{\varepsilon _{a}+\varepsilon _{e}}{2}+\frac{t_{ae}}{2}%
\cos \frac{ak_{y}}{\sqrt{3}}+\sqrt{3}t_{ae}\sin \frac{ak_{y}}{\sqrt{3}}, 
\notag \\
F_{bb}^{M}& =\frac{\varepsilon _{b}+\varepsilon _{d}}{2}+\frac{t_{bd}}{2}%
\cos \frac{ak_{y}}{\sqrt{3}}+\sqrt{3}t_{bd}\sin \frac{ak_{y}}{\sqrt{3}}, 
\notag \\
F_{cc}^{M}& =\frac{\varepsilon _{a}+4\varepsilon _{c}+\varepsilon _{e}}{6}+%
\frac{2t_{ac}-t_{ae}+2t_{ce}}{6}\left( \cos \frac{ak_{y}}{\sqrt{3}}+\sqrt{3}%
\sin \frac{ak_{y}}{\sqrt{3}}\right) ,  \notag \\
F_{ab}^{M}& =\left( \frac{e^{-2i\pi /3}}{4}t_{ab}e^{i\frac{ak_{y}}{2\sqrt{3}}%
}+t_{de}e^{-i\frac{ak_{y}}{2\sqrt{3}}}\right) \cos \frac{k_{x}}{2},  \notag
\\
F_{ab}^{M}& =\frac{e^{-i\pi /3}\left( 2t_{ac}-t_{ae}\right) e^{-i\frac{ak_{y}%
}{\sqrt{3}}}+\left( t_{ae}+t_{ce}\right) e^{-i\frac{ak_{y}}{\sqrt{3}}}}{2%
\sqrt{3}}+\frac{-e^{i\pi /3}\left( \varepsilon _{a}-\varepsilon _{e}\right) 
}{2\sqrt{3}},  \notag \\
F_{bc}^{M}& =\left( \left( t_{ab}+2t_{cd}\right) e^{-i\frac{ak_{y}}{2\sqrt{3}%
}}+\frac{\left( -1\right) ^{1/3}}{\sqrt{3}}\left( t_{bc}+2t_{de}\right) e^{i%
\frac{ak_{y}}{2\sqrt{3}}}\right) \cos \frac{ak_{x}}{2}.
\end{align}%
\end{widetext}

\end{document}